\documentclass{article}

\usepackage{graphicx}
\usepackage{amssymb}
\usepackage{slashed}
\usepackage{amssymb}
\usepackage{amsmath}
\usepackage{amsfonts}
\usepackage{latexsym}
\usepackage{slashed}
\usepackage{slashbox}
\usepackage{url}
\usepackage{float}
\usepackage{subfig}
\begin{document}
\title{Application of Jet Trimming in Boosted Higgs Search\footnote{Talk presented by Wenhan Zhu at Pheno 2011, Madison, Wisconsin, May 2011.}}
\author{Wenhan Zhu \\
Department of Physics, Princeton University, Princeton, NJ 08544} 
\date{}
\maketitle
\begin{abstract}
We present the study of the $WH$ and $ZH$ search with the Higgs Boson decayed to $b\bar{b}$ at the Large Hadron Collider. The Higgs Boson and the Vector Boson are required to be boosted, and the Higgs Boson is reconstructed with Jet Trimming Technique. The statistical significance for 30$fb^{-1}$ data is 4.5 $\sigma$, which is comparable to the previous result~\cite{Butterworth:2008iy}. 
\end{abstract}

\section{Introduction}
The Higgs Boson search is the most important search at the Large Hadron Collider (LHC), because it is an essential part of the standard-model electroweak symmetry breaking. Current electroweak fits, together with the LEP and Tevatron exclusion limit, favour a light Higgs boson one with mass around 120 GeV~\cite{Grunewald:2007pm}~\cite{Aaltonen:2010yv}. It is a challenging task for the discovery of  Higgs Boson at this mass region~\cite{Aad:2009wy}~\cite{Ball:2007zza}.

The Mass Dropping and Filtering method introduced in~\cite{Butterworth:2008iy} makes the Higgs Boson production channel associated with a vector boson very promising. In this channel, the Higgs Boson will decay hadronically into two b-tagged jet with the vector boson decay leptonically. The dominant background for this process is $VV$,$Vj$ and $t\bar{t}$. We will employ a similar kinematic selection to~\cite{Butterworth:2008iy}, but we will reconstruct the Higgs Boson using the jet trimming technique~\cite{Krohn:2009th}.

\section{Jet Trimming}
Jet Trimming~\cite{Krohn:2009th} is a designated procedure for removing the ISR/MI/Pileup from the FSR. The intrinsic idea is that ISR/MI/Pileup will be much softer than the FSR, therefore, we will form a fat jet using a larger cone and then recluster the fat jet with a smaller cone and throw away the softer subjets. 

We will make some change to the original algorithm described in~\cite{Krohn:2009th}  for the jet substructure of the boosted Higgs Boson. First, we will find the two b-tagged jets by clustering the jet constituents of the fat jet. Also, the Higgs jet, different from a QCD jet, is a dipole itself, so we expect to have more radiation between the two b quarks. Therefore, we will use a dynamical $f_{\rm cut}$, which is proportional to the distance between the subjet and the fat jet. 

The jet trimming algorithm proceeds as follows:
\begin{itemize}
\item Cluster all the final state particles with Fastjet 2.4.2~\cite{Cacciari:2005hq},anti$k_T$ algorithm with a cone size 1.2.
\item Cluster the particles in the hardest jet with a smaller cone size 0.3 with anti$k_T$ algorithm to find out the hardest two subjets, and we will require each of the two subjets to be b tagged. We assume a 60\% tagging efficiency and 2\% of mistagging efficiency.
\item Cluster the remaining particles with an even smaller cone size 0.2 with $k_T$ algorithm to form the subjets.
\item If $p_{T}^{i} > f_{\rm cut} p_{T} \Delta R$ the subjet is kept else it is trimmed, $p_{T}$ is the $p_T$  of the fat jet and   $\Delta R$ is the distance between the subjet and the fat jet. The $f_{\rm cut}$ is chosen to be 0.03 in this analysis.
\item Now we have the Higgs Candidate. We will require the Higgs Candidate $p_T$ larger than 200 GeV and $\eta$ less than 2.5.
\end{itemize}

\section{Results}
The events are generated by Pythia 6.403~\cite{Sjostrand:2006za}, fully showered and hadronized. The underlying event is incorporated by Pythia "DW" tune.  For this analysis, signal samples of $WH, ZH$ were generated, as well as $WW,ZW,ZZ,Z+{\rm jet},W+{\rm jet},t\bar{t}$ to study backgrounds.

There are three search channels in this analysis and the channel specific cuts are very similar to~\cite{Butterworth:2008iy}:
\begin{itemize}
\item Leptonic channel: two opposite sign lepton ($e$ or $\mu$) with $p_{T}>30$ GeV and $|\eta|<2.5$, with an invariant mass between 80 and 100 GeV.
\item Missing $E_T$ channel: Missing Transverse momentum$>200$GeV.
\item Semi-leptonic Channel: Missing transverse momentum$>30$GeV plus a lepton ($e$ or $\mu$) with $p_T>30$GeV. Veto event if there is jet with  $p_{T}>30$ GeV and $|\eta|<3.0$.
\item all channel: no more lepton with $p_{T}>30$ GeV and $|\eta|<2.5$ except to reconstruct the vector boson, no more b-tagged jets with $p_{T}>30$ and $|\eta|<2.5$. 
\end{itemize}

\begin{figure}[t]
\includegraphics[scale=0.35,angle=90]{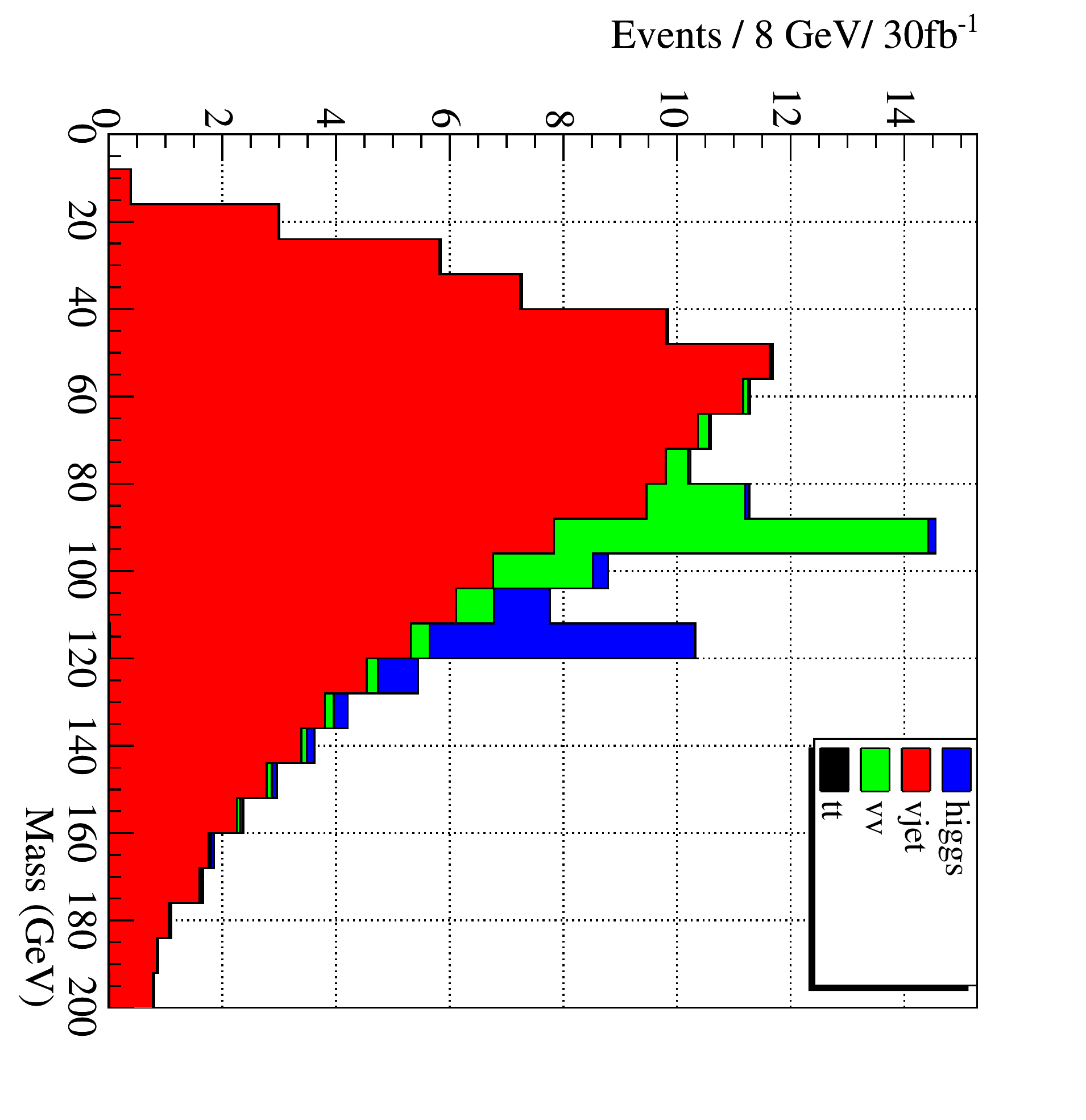}
\includegraphics[scale=0.35,angle=90]{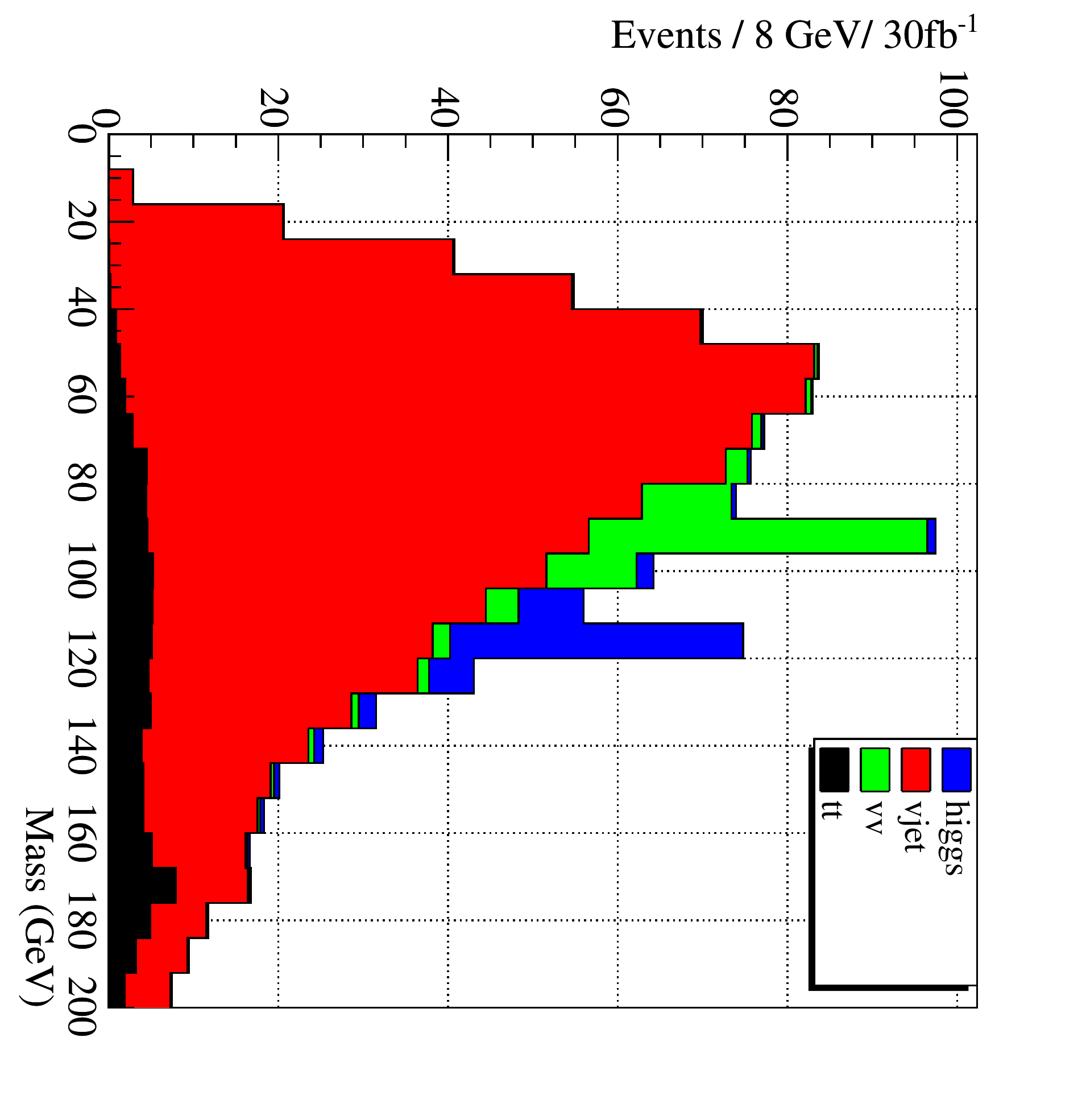}
\includegraphics[scale=0.35,angle=90]{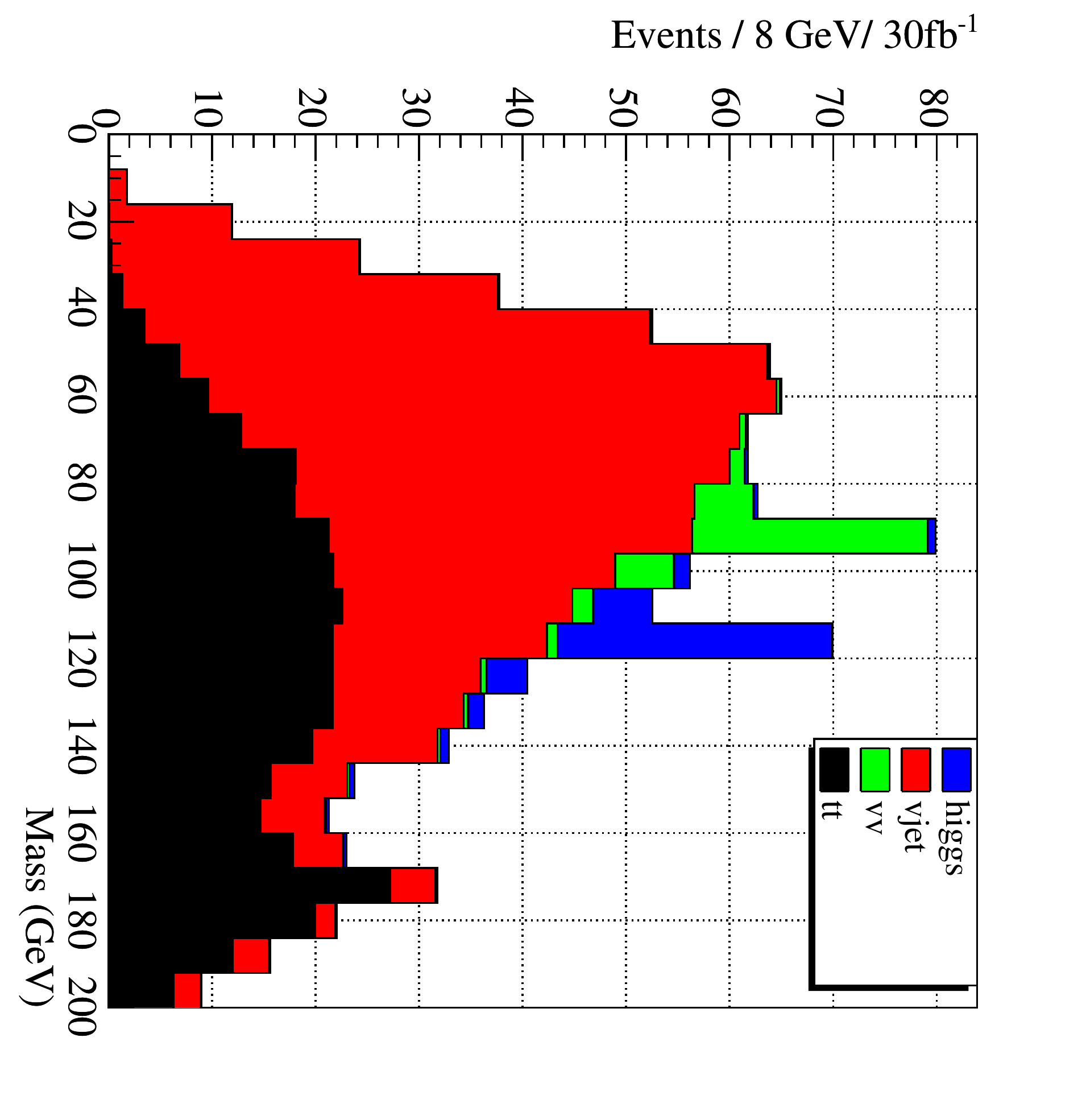}
\includegraphics[scale=0.35,angle=90]{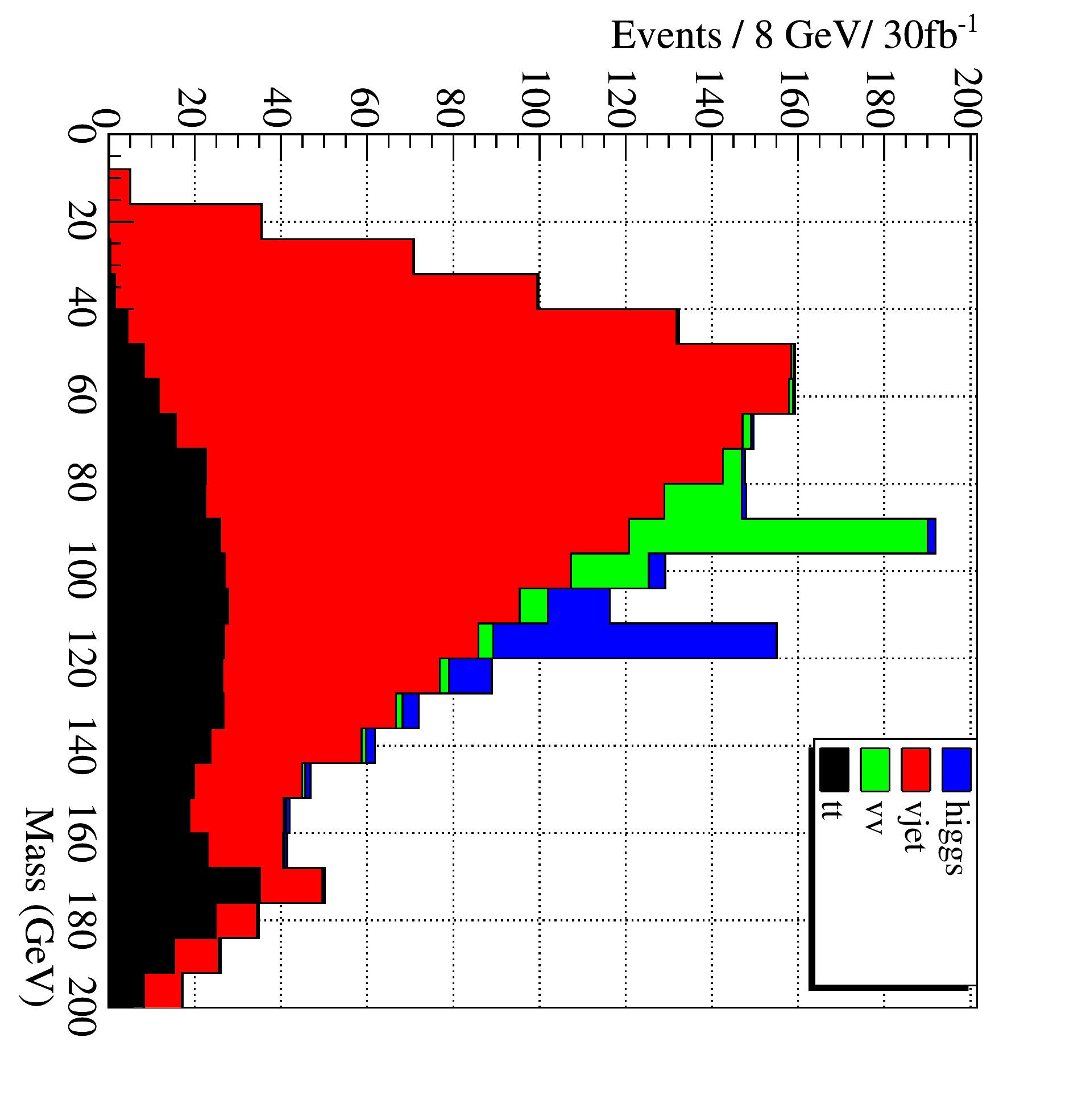}
\caption{The Signal and Background for a 115 GeV SM Higgs Boson for 30 $fb^{-1}$. On the top, the left plot is or leptonic channel and the right plot is for missing Et channel. On the bottom, the left plot is for semi-leptonic channel and the right plot is the total signal and background for all the channel.  }
\label{fig:mass}
\end{figure}

The mass spectrum of the Higgs Candidate with $m_H=$115 GeV is shown in Fig~\ref{fig:mass} for the three sub-channel and combined channel. The number of both signal and background for Higgs Mass between 112-128 GeV for 30$fb^{-1}$ data is shown in Table~\ref{tab:all}, the significance is 4.5 $\sigma$(8.2$\sigma$ for 100 $fb^{-1}$). The result is comparable to ~\cite{Butterworth:2008iy} and offer an alternative strategy for search for boosted Higgs. 
\begin{table}[t]
\parbox{\textwidth}{
\begin{center}
\begin{tabular} {|c|c|c|c|c|c|c|}
\hline
Channel    & Signal & V+Jet & $t\bar{t}$ & VV & S/B & S/$\sqrt{B}$\\
\hline 
Leptonic &5.4 &10.0  &0.032 &0.53  &0.51 &1.66   \\
Missing $E_t$ &  24.3& 65.6 &12.9& 3.4  &0.3& 2.7 \\
Semi-leptonic &30.6   & 35.0   & 49.9  &1.6 &0.35 &3.3 \\
Total &60.3  &110.60  &62.8  &6.5 &0.34  &4.5  \\
\hline
\end{tabular}
\end{center}
\caption{\label{tab:all} Signal and Background for a 115 GeV SM Higgs Boson for 30 $fb^{-1}$. }
}
\end{table}

\section{Conclusion and Outlook}
Here we have applied jet trimming technique to the boosted Higgs Boson search for a low mass(115 GeV) SM Higgs Boson. The statistical significance for 30$fb^{-1}$ data is 4.5 $\sigma$, which is comparable to the previous result ~\cite{Butterworth:2008iy}. This could be considered as an alternative search strategy for the high-$p_T$ $WH$,$ZH$ channel at the LHC. 

\section{Acknowledgement}
The author would like to acknowledge useful discussions with Valerie Halyo, David Krohn, Gavin Salam and Lian-tao Wang.

\end{document}